\documentstyle[twoside,fleqn,espcrc2,epsf]{article}

\newcommand{\la}{\raise.16ex\hbox{$\langle$}}
\newcommand{\ra}{\raise.16ex\hbox{$\rangle$}}
\newcommand{\be}{\begin{equation}}
\newcommand{\ee}{\end{equation}}
\newcommand{\bea}{\begin{eqnarray}}
\newcommand{\eea}{\end{eqnarray}}

\newcommand{\gl}{\mathrel{\rlap{\lower2pt\hbox{\small \hskip1pt$<$}}
\hspace*{1pt}\raise3pt\hbox{\small $>$}}}

\title{'t Hooft loops and consistent order parameters for confinement}

\author{
Ph.~de~Forcrand~$^{a,b}$
\thanks{Talk presented by Ph.~de~Forcrand}
and  
L.~von~Smekal~$^c$\\
\vspace{3mm}
$^a$~Institut f\"ur Theoretische Physik, ETH-H\"onggerberg, CH-8093 Z\"urich,
Switzerland\\
$^b$~Theory Division, CERN, CH-1211 Geneva 23, Switzerland\\
$^c$~Institut f\"ur Theoretische Physik III, Universit\"at
  Erlangen-N\"urnberg, D-91058 Erlangen, Germany
}

\begin{document}

\markright{\small\sf CERN-TH/2001-281, FAU-TP3-01/9,  hep-lat/0110135}

\begin{abstract}
We study ratios of partition functions in two types of sectors of $SU(2)$, 
with fixed temporal center flux and with static fundamental charge.
Both can be used as bona fide order parameters 
for the deconfinement transition.
\end{abstract}

\maketitle

\section{Introduction}
It is well known that the deconfinement transition which occurs at a critical
temperature $T_c$ in Yang-Mills theory corresponds to the spontaneous 
breaking of center symmetry. It is natural then to expect that the center
degrees of freedom of the gauge field are somehow responsible for 
confinement and its disappearance.
Indeed, numerical studies tend to show that the center vortices, extracted
from the gauge field after partial gauge-fixing, confine linearly at
$T \sim 0$ with the Y-M string tension \cite{Greensite}, 
at least in the continuum limit \cite{MP1}.
Here, we show quantitatively, in a gauge-invariant way, that the free
energy of a center vortex is an order parameter for the deconfinement
transition, as anticipated in \cite{KT}.
It jumps from $0$ to $+\infty$ as the temperature is raised. 

Alternatively, one should be able to use the free energy $F_q$ of a 
static fundamental charge, jumping from $+\infty$ to $0$ at $T_c$.
The Polyakov loop $P$ is commonly used for this purpose in lattice studies.
If $\langle Tr P \rangle = e^{-\frac{1}{T} F_q}$, the symmetric (broken)
phase gives for $F_q$ an infinite (finite) value. However, the periodic
boundary conditions (b.c.) within which $\langle Tr P \rangle$ is measured
are incompatible with the presence of a single charge. And, like any
Wilson loop, $\langle Tr P \rangle$ is subject to UV-divergent
perimeter terms, such that $\langle Tr P \rangle = 0$ at all temperatures
as the lattice spacing $a \rightarrow 0$.
Here, following \cite{tHooft}, we measure the gauge-invariant, UV-regular
free energy of a static fundamental charge, and show that it has the
expected behaviour, dual to that of a center vortex.
Moreover, the string tension below $T_c$ and the dual (vortex) string
tension above $T_c$ \cite{PRL} are quantitatively related by a universal
ratio, for more details, see \cite{full}.

\section{'t Hooft loop, center vortices, and t.b.c.}

The 't Hooft loop $\tilde W$, defined on the dual lattice, multiplies 
every Wilson loop linked to it by a $Z_N$ center element. 
In the case of an $R_x \times R_y$ loop in $SU(2)$, it is constructed 
by flipping the sign of the coupling $\beta \rightarrow -\beta$ of the 
$R_x \times R_y$ $(z,t)$-plaquettes dual to $\Sigma$, 
the rectangle bounded by $\tilde W$.
[To compute the 't Hooft loop, we actually flip $\beta$ one plaquette 
at a time as in \cite{PRL}.]
Equivalently, as shown explicitly by a change of integration variables,
one could choose for $\Sigma$ any other surface bounded by $\tilde W$.
In any case, a set of plaquettes will carry the extra $Z_N$ flux of a 
center vortex, whose precise location is gauge-dependent.
The gauge-invariant 't Hooft loop $\tilde W$ therefore generates a fluctuating
center vortex sheet bounded by $\tilde W$. \\
When an $L \times L$ spatial 't Hooft loop is created in an $L^3 \times N_t$
lattice, every $L \times N_t$ temporal plane dual to it must carry a
$Z_N$ flux. This is equivalent to enforcing twisted boundary conditions
(t.b.c.) in this temporal plane. 
Calling $Z_0$ the periodic b.c. partition function,
we measure the free energy cost $e^{-F_k} = Z_k(i)/Z_0$ of t.b.c. in 
$i=1, 2$ and 3 temporal planes as a function of temperature $T$.
Our results for $Z_k(1)/Z_0$ are shown in Fig.~1 for 
lattices of size $N_t=4, L=6..20$.
As expected from Finite-Size Scaling (FSS), they are well described
around $T_c$
by a single function of $L/\xi$, where the correlation length $\xi$ diverges 
near $T_c$ as $|t|^{-\nu}$, with $t = \frac{T}{T_c}-1$ the reduced 
temperature and $\nu \approx 0.63$ for the $3d$ Ising model universality 
class. This is shown in Fig.~2, where the FSS variable is 
$x =  L T_c |t|^\nu$. The large-$x$ asymptotic behaviour shows
an area law $exp(-\tilde\sigma_0 x^2)$ in the deconfined phase, and
screening $exp(exp(-b x))$ in the confined phase. 
In the thermodynamic limit, $Z_k/Z_0$ is a step function and can be
used as a UV-regular order parameter as advertised.
The interpretation of this behaviour is simple.
Twisted b.c., or equivalently large center vortices, can be imposed
at exponentially small cost when the system size becomes large compared to
the confinement correlation length at $T\sim 0$. 
At finite temperature, the spread of the center vortex in a temporal plane
is now limited by the compact dimension $1/T$. Above $T_c$, the Polyakov
loop acquires a non-zero expectation value. Temporally twisted b.c.,
which enforce antiperiodicity on the Polyakov loop, then create 
an interface of minimal area $L^2$ separating positive and negative
Polyakov loops. Such interfaces and the corresponding center vortices
are exponentially suppressed by the dual-string or interface tension. 
T.b.c.~in 2 and 3 temporal planes modify the minimal area of the
interface to $L^2 \sqrt{2}$ and $L^2 \sqrt{3}$ respectively. We observe
that the dual string tension $\tilde\sigma_0$ is multiplied by precisely
$\sqrt{2}$ and $\sqrt{3}$.

\section{Confinement/screening of fundamental charge: electric flux}

As seen above, temporally (say, $(z,t)$) twisted b.c. enforce antiperiodicity
on the Polyakov loop: $Tr P^\dagger(\vec{x}) P(\vec{x}+L\hat{z}) = -1$,
instead of $+1$ with periodic b.c. Consider then the expectation value
of this binary observable, in an enlarged partition function $Z_{enl}$
which sums over all possible temporal b.c., periodic or twisted in each
plane. $Z_{enl} = Z_0 + 3 Z_k(1) + 3 Z_k(2) + Z_k(3)$.
Contrary to $Z_0$, $Z_{enl}$ allows for the consistent introduction of
a static fundamental charge, which is neutralized by its image in the
twisted direction (say $\hat z$). Keeping track of the $(z,t)$-flux for
each b.c. \cite{full}, 
\bea
&&\langle Tr P^\dagger(\vec{x}) P(\vec{x}+L\hat{z}) \rangle
\equiv Z_e(1)  \\
&&~~~~~~= \frac{1}{Z_{enl}} (Z_0 + Z_k(1) - Z_k(2) - Z_k(3))
\nonumber
\eea
Similar expressions (in fact, $Z_2$ Fourier transforms of the
$Z_k$'s) follow for 2 and 3 orthogonal 'electric' fluxes \cite{tHooft,H2N}.
Fig.~3 shows $Z_e(1)$. Its behaviour is dual to that of $Z_k(1)$.
It also is a function of the FSS variable $x$, and it shows screening
$exp(exp(-\tilde b x^2))$ above $T_c$ and confinement $exp(-\sigma_0 x)$
below $T_c$, see Fig.~4.
Moreover, because $Z_e$ and $Z_k$ are simply related by Fourier transform,
one obtains that $\tilde b = \tilde\sigma_0$ and $b=\sigma_0$.

Universality states that the FSS functions measured here for $SU(2)$ 
are the same as for the $3d$ Ising model after a single rescaling of $x$.
This allows a precise determination of the large-$x$
coefficients $\sigma_0, \tilde\sigma_0$ by matching only the small-$x$
part of the FSS functions between the two models \cite{MP2}.
Another universal ratio is given by
\begin{equation} 
\tilde\sigma_0/\sigma_0^2 = 
\tilde\sigma(T_c^+)/\left( \frac{\sigma(T_c^-)}{T_c^-} \right)^2 \approx 0.41
\; .
\end{equation}
Our data is consistent with this prediction, which
relates the two sides of the phase transition.

\begin{figure}[t]
\vspace*{-.5cm}
\epsfxsize=8truecm
\epsfysize=6.0truecm
\hskip -.55cm\mbox{\epsfbox{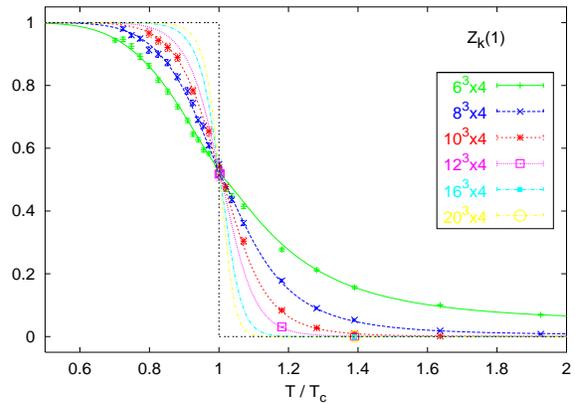}}
\vskip -1.2cm
\caption{Partition function of one temporal twist as a function of
temperature for various volumes.}
\vskip -0.5cm
\label{ZkT} 
\end{figure}

\begin{figure}[t]
\vspace*{-.5cm}
\epsfxsize=8truecm
\epsfysize=6.0truecm
\hskip -.55cm\mbox{\epsfbox{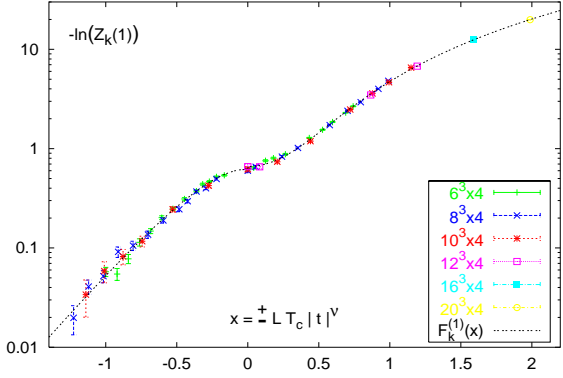}}
\vskip -1.2cm
\caption{Free energy of one temporal twist 
as a function of the FSS variable $x\gl 0$ 
for $T\gl T_c$.}   \label{tH1all} 
\vskip -0.5cm
\end{figure}

\begin{figure}[t]
\vspace*{-.5cm}
\epsfxsize=8truecm
\epsfysize=6.0truecm
\hskip -.55cm\mbox{\epsfbox{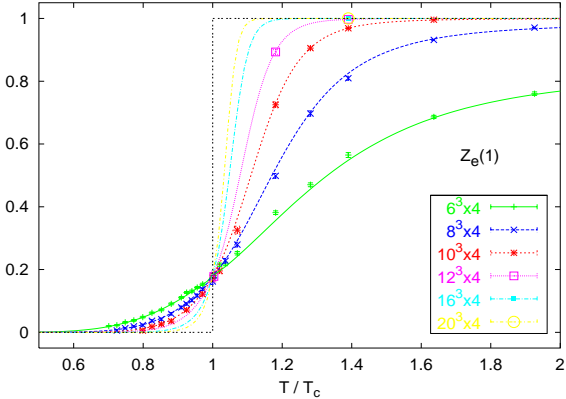}}
\vskip -1.2cm
\caption{Partition function of one electric flux.}
\vskip -0.5cm
\label{ZeT} 
\end{figure}

\begin{figure}[t]
\vspace*{-.5cm}
\epsfxsize=8truecm
\epsfysize=6.0truecm
\hskip -.8cm\mbox{\epsfbox{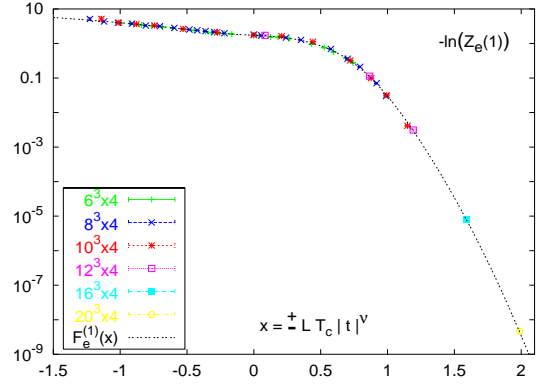}}
\vskip -1.2cm
\caption{Free energy of one electric flux as a function of the FSS variable 
$x \gl 0 $ for $T \gl T_c$.} 
\vskip -0.5cm
 \label{E1} 
\end{figure}

\section{Summary}

Let us give an intuitive description of the deconfinement transition
in Yang-Mills theory: \\
$\bullet$ At low temperature, center vortices can spread to lower their
free energy. Their proliferation disorders the Wilson loop and makes the
average Polyakov loop traceless, reflecting confinement. \\
$\bullet$ As the temperature is raised, the squeezing of center vortices in 
the compact $1/T$ dimension prevents them from arbitrarily lowering their free
energy. Eventually, they cannot disorder the Polyakov loop anymore.
Macroscopic regions of Polyakov loops of a definite center sector appear,
separated by interfaces whose tension exponentially suppresses center vortices.
Confinement has disappeared. This squeezing does not affect center flux in
spatial planes, so that the spatial string tension persists at all
temperatures. 

Because the free energy $F_q$ of a fundamental charge results from the
cancellation of partition functions with different b.c., $Z_e =
e^{-\frac{1}{T} F_q}$ is 
exponentially small in the confined phase, and $\sim 1$ in the deconfined
phase. We have shown that the ratios of partition functions $Z_k/Z_0$ or
$Z_e/Z_0$, which are well-defined continuum quantities, can both be used as
order parameters for the transition.

If dynamical fundamental charges (quarks) are introduced in the system,
the picture becomes murky, reflecting the change from a phase transition
to a crossover. To start with, the 't Hooft loop now depends on the 
particular surface $\Sigma$ along which center flux is introduced. 
Its relevance in full QCD remains to be studied. \\

We thank M.~Garc\'{\i}a~P\'erez, M.~Hasenbusch and M.~Pepe for discussions.

\markright{}

\end{document}